\documentclass[12pt]{jpconf}

\usepackage{citesort}
\usepackage{graphicx}

\begin{document}
	\nocite{*}
	
	\title{Induced superconducting pair correlations in a quasicrystal coupled to a BCS superconductor}
	
	\author{ Gautam Rai$^1$, Stephan Haas$^1$ and A~Jagannathan$^2$}
	
	\address{$^1$ Department of Physics and Astronomy, University of Southern California, Los Angeles, California 90089-0484, USA}
	\address{$^2$Laboratoire de Physique des Solides, Universit\'e Paris-Sud, 91405 Orsay, France}
	
	\section{Introduction}
	
	The term proximity effect refers to the superconducting order induced within a normal conductor (N) when it is placed in contact with a superconductor (S). This effect was studied in the early 1960s by de Gennes and coworkers who showed for example that the induced order parameter $\Delta$ in a disordered normal metal decays exponentially as a function of the distance $d$ from the N-S interface, that is, $\Delta \sim \exp^{-d/\xi}$, where the penetration depth $\xi$ depends on the diffusion coefficient and on the temperature. For low enough temperatures, therefore, one can induce superconductivity throughout the sample for sufficiently small widths. The proximity effect has been revisited recently in a large variety of new situations in order to probe novel electronic states, as in the case of the Dirac electrons in graphene or Majorana fermions in topological insulators. It is therefore interesting to ask what the nature of the proximity effect which arises when a quasicrystal is placed in contact with a superconductor. Recall that these materials possess conduction electrons and have a Fermi surface, and yet are extremely poor conductors. This is due to the spatial structure of electronic states which are typically neither fully extended nor localized but in some sense ``in-between". 
	
	To gain some insight into the nature of the proximity effect in this new context, we consider the simplest known model of a quasicrystal,  the 1D Fibonacci chain. For the tight-binding hopping model on this structure all states are known to be critical and their multifractal exponents have been computed  \cite{mace2016fractal}. We couple a Fibonacci approximant chain to a BCS-type superconductor, so as to form a hybrid ring. The local superconducting pairing order parameters of this system are calculated \`a la Bogoliubov-de Gennes (BdG), a mean field approach, which can be justified due to the fact that our hybrid ring is in fact imbedded in a 3D system. By solving the resulting BdG equations self-consistently, we show that the proximity effect in the Fibonacci chain is long ranged, i.e. decays as a power law as one goes away from the N-S interface. We show that the order parameter can be larger than in periodic chains, due to the large fluctuations inherent to critical states. Finally, we show that the induced superconducting order carries information on topological properties, namely, the winding number of edge states of the Fibonacci chain.  
	
	\begin{figure}[h]
		\centering
		\includegraphics[scale=0.4]{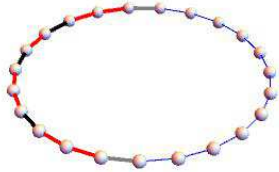} 
		\caption{Hybrid ring consisting of a Fibonacci chain with two different hopping amplitudes $t_A$ and $t_B$ (represented by thick red and black bonds) and a periodic superconducting chain (thin blue bonds). The interface hopping amplitude is represented by a thick grey bond.}
		\label{hist.fig}
	\end{figure}

	\section{The hybrid NS ring}
	
	In our tight-binding model the N component of the ring is an approximant of the Fibonacci chain with two hopping amplitudes $t_A$ and $t_B$ and the S component is a periodic $s$-wave superconductor. The two are coupled so as to form a ring, as shown in Fig.1, where the Fibonacci chain is shown with thick red and black bonds, and the periodic superconducting chain is shown with thin blue bonds.  We assume a particle-hole symmetric case with half-filling, so that $E_F=0$ in both the normal and superconducting parts.  We now give the ingredients that go into building our model.\\
	
	\noindent
	{\sl{In the normal part}} the hopping amplitudes are either $t_A$ or $t_B$, arranged in a Fibonacci sequence.   
	The number of sites is $N=35,56,90,....$ -- i.e. the length of approximant chains plus 1. 
	The Hamiltonian for the FC depends implicitly on an additional parameter, $\phi$, which determines the sequence of hopping amplitudes for a given length \cite{preprintproximity}. This parameter, termed ``phason angle" in analogy with the Aubry-Andr\'e model for incommensurate charge density wave systems \cite{aubry1980analyticity} can be varied between 0 and $2\pi$ and in this manner one can generate the complete set of approximants of length $N$. Successive chains in this set only differ by a single local phason flip, and the total number of distinct chains obtained by varying $\phi$ is exactly $N$. Taking  open boundary conditions and solving for the energy spectra for each value of $\phi$, it can be seen that as $\phi$ is varied, the energies of certain states move within the gaps of the spectrum. The number of times that a given state traverses the gap corresponds, precisely, to the label of the gap in which the state lives \cite{bellissard1989spectral}. Fig.2 shows the integrated DOS for an approximant chain versus the energy, along with gap labels of some of the principal gaps (as the spectrum is symmetric, only the lower half of the spectrum is shown). This phenomenon was first shown, in \cite{baboux2017measuring} by direct experimental measurements of Fibonacci approximant chains using polaritonic cavity modes.
	\begin{figure}[h]
		\centering
		\includegraphics[scale=0.5]{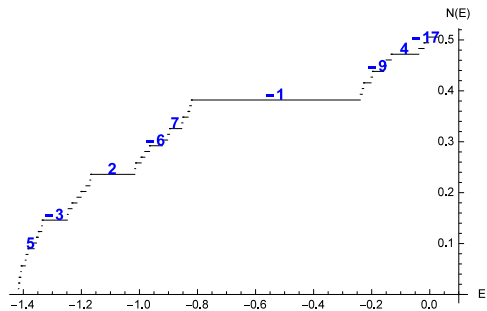} 
		\caption{ Integrated density of states (IDOS) for a chain of 89 sites showing the main gaps and their associated labels}
		\label{deltactr.fig}
	\end{figure}
	
	\noindent
	{\sl{ In the superconducting part}} we assume a constant hopping amplitude and attractive Hubbard interactions of strength $V$ on each site. We perform a mean field decoupling of the Hubbard interaction terms, by introducing a set of (real) order parameters $\Delta_i=\langle c_{i\downarrow}c_{i\uparrow}\rangle$ for each site. The full mean field Hamiltonian for the hybrid ring then reads
	\begin{eqnarray}
	H =& \sum_{i,\sigma} \left\{-t_i (c^\dagger_{i+1,\sigma} c_{i,\sigma} + c^\dagger_{i,\sigma} c_{i+1,\sigma}) + U_i c^\dagger_{i,\sigma} c_{i,\sigma} \right\}  
	- \sum_{i} V_i \left( c^\dagger_{i\uparrow} c^\dagger_{i_\downarrow}+c_{i\uparrow} c_{i\downarrow} \right)
	\end{eqnarray}
	where the sums run over two spin states $\sigma=\pm 1$, and over sites $i=1$ to $N$ for the quasicrystalline part and from $N+1$ to $N+N_s$ in the superconducting part. The site-dependent mean field onsite energy terms $U_i$, and off-diagonal terms $V_i=V\Delta_i$ are generated by the interactions in the superconductor.  
	
	The parameters of the model are chosen as follows:  We fix the values of the hopping in the periodic chain and across the interfaces (assumed to be transparent) to be equal to $t_B$ (which fixes the energy scale). We take an interaction $V=0.8$ of intermediate strength. The remaining variables are $\phi$, which determines the FC structure, and $t_A/t_B$, the hopping ratio. This last parameter allows to go from strongly modulated quasiperiodicity (when it takes very small values) all the way upto the periodic chain (when it has the value 1). With these assumptions, the resulting quadratic Hamiltonian was solved by self-consistent numerical calculation and for temperature $T=0$.   
	\section{Solutions for the superconducting order in the FC}
	Fig.3 shows the solution for the pair amplitudes as a function of the site index for a single hybrid ring (with $N=N_s=90$) and for a value of the hopping ratio $t_A/t_B=0.9$. The pair amplitude in the quasicrystal decays as one goes from the interfaces (situated on sites 1 and N) towards the center, albeit with strong fluctuations from site to site. The values of the pair amplitudes depend on the chain length, on the hopping ratio, and lastly on the phason angle. We now focus on the value at the center of the chain, namely, $\Delta_{mid}$, in order to study the strength of penetration of the superconducting correlations as a function of the phason angle. To extract this quantity, we first fit the data to a power law, and then evaluated the fit function at the center of chain. This quantity $\Delta_{mid}$ is independent of the fitting procedure and gives a more reliable measure of the strength of the penetration than the decay exponent which is harder to obtain by a fit. Fig.4 shows $\Delta_{mid}$ plotted versus $\phi$, for several chains.
	
	\begin{figure}[h]
		\centering
		\includegraphics[scale=0.28]{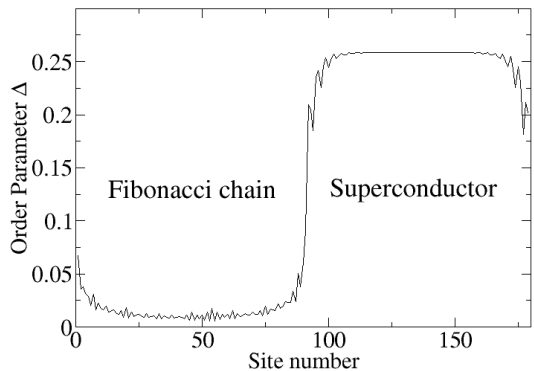} 
		\caption{Pair amplitudes as a function of site index for a hybrid chain of total length 180 sites and for $t_A/t_B=0.9$ }
		\label{op.fig}
	\end{figure}
	
	\begin{figure}[h]
		\centering
		\includegraphics[scale=0.35]{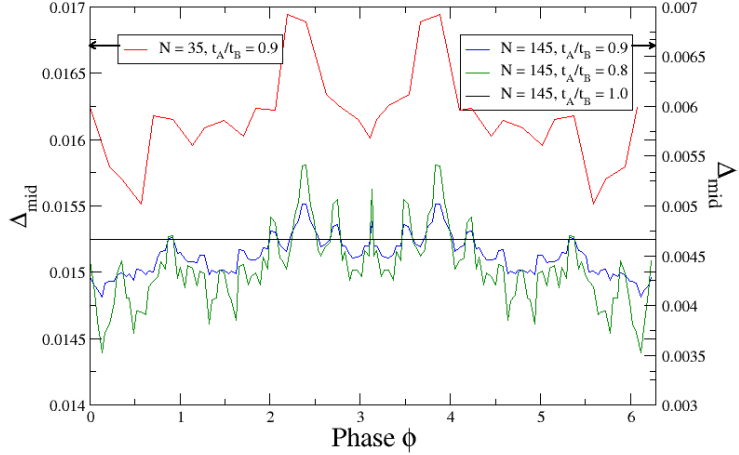} 
		\caption{ Pair amplitude at the midpoint $\Delta_{mid}$ as a function of phason angle $\phi$ for two different system sizes (N=35 and 144) and for three different values of the hopping ratio.}
		\label{deltactr.fig}
	\end{figure}
	
	As Fig.4 shows, $\Delta_{mid}$ versus $\phi$ has a complex structure of peaks and valleys. A fourier analysis of this function yields periodicities of $4, 17, 9, 21, 6$ in order of decreasing strengths(see \cite{preprintproximity} for further details of the computations).  These periodicities correspond to the absolute value of gap labels of Fig.2, in particular of gaps which are largest and/or closest to the Fermi level $E_F=0$. The explanation : as stated before, the gap label is a topological number describing the winding number of the states in that gap. As the phason angle is varied, the edge states can become delocalized within the chain, resulting in the peaks and valleys of $\Delta_{mid}$. Compared to a periodic chain, $\Delta_i$ can be quite large due to local fluctuations. If however we consider an averaged quantity -- the pair amplitude on each site after averaging over all values of $\phi$, fluctuations are smoothed out. This quantity can be fitted to a power law decay in the distance $i$ from the interface, i.e. $\overline{\Delta_i} \sim i^{-\alpha}$, where the power $\alpha$ is non-universal and depends on the hopping ratio. A more detailed analysis of the power law dependence is in progress and will be reported elsewhere.

	\section{Discussion and conclusions}
	We have shown that the proximity effect in Fibonacci chain approximants is long ranged, typically decaying as a power law. Due to the large fluctuations of critical states, the pair amplitude can be larger than in periodic chains. The proximity induced order parameters strongly depend on the topological edge states present in the FC. If one cycles through the family of approximant chains, corresponding to different values of the phason angle, the order parameter in the midpoint of the chain can be seen to follow the winding numbers of the edge states. A similar calculation could in principle be carried out for higher dimensions, and show a long range (power law) proximity effect. These predictions could be observable in experiments on quasiperiodic thin films as for example in  \cite{ledieu2014surfaces} in which the proximity effect could be induced via superconducting contacts and studied by scanning tunneling spectroscopy.

	\section*{References}
	\bibliography{iobib}
	
\end{document}